\documentclass[aps,pra, twocolumn, amsmath, amssymb, showpacs, superscriptaddress]{revtex4-2}

\usepackage{graphicx}
\usepackage{bm}

\usepackage{xcolor}

\begin{document}

\title{Lineshape-asymmetry-caused shift in atomic interferometers}

\author{V.\,I.\,Yudin}
\email{viyudin@mail.ru}
\affiliation{Novosibirsk State University, ul. Pirogova 1, Novosibirsk, 630090, Russia}
\affiliation{Institute of Laser Physics SB RAS, pr. Akademika Lavrent'eva 15 B, Novosibirsk, 630090, Russia}
\affiliation{Novosibirsk State Technical University, pr. Karla Marksa 20, Novosibirsk, 630073, Russia}
\author{O.\,N.\,Prudnikov}
\affiliation{Novosibirsk State University, ul. Pirogova 1, Novosibirsk, 630090, Russia}
\affiliation{Institute of Laser Physics SB RAS, pr. Akademika Lavrent'eva 15 B, Novosibirsk, 630090, Russia}
\author{A.\,V.\,Taichenachev}
\affiliation{Novosibirsk State University, ul. Pirogova 1, Novosibirsk, 630090, Russia}
\affiliation{Institute of Laser Physics SB RAS, pr. Akademika Lavrent'eva 15 B, Novosibirsk, 630090, Russia}
\author{M.\,Yu.\,Basalaev}
\affiliation{Novosibirsk State University, ul. Pirogova 1, Novosibirsk, 630090, Russia}
\affiliation{Institute of Laser Physics SB RAS, pr. Akademika Lavrent'eva 15 B, Novosibirsk, 630090, Russia}
\affiliation{Novosibirsk State Technical University, pr. Karla Marksa 20, Novosibirsk, 630073, Russia}
\author{D.\,N.\,Kapusta}
\affiliation{Institute of Laser Physics SB RAS, pr. Akademika Lavrent'eva 15 B, Novosibirsk, 630090, Russia}
\author{A.\,N.\,Goncharov}
\affiliation{Institute of Laser Physics SB RAS, pr. Akademika Lavrent'eva 15 B, Novosibirsk, 630090, Russia}
\author{M.\,D.\,Radchenko}
\affiliation{Novosibirsk State University, ul. Pirogova 1, Novosibirsk, 630090, Russia}
\affiliation{Institute of Laser Physics SB RAS, pr. Akademika Lavrent'eva 15 B, Novosibirsk, 630090, Russia}
\affiliation{Novosibirsk State Technical University, pr. Karla Marksa 20, Novosibirsk, 630073, Russia}
\author{V.\,G.\,Pal'chikov}
\affiliation{All-Russian Research Institute of Physical and Radio Engineering Measurements, Mendeleevo, Moscow region, 141570 Russia}
\affiliation{National Research Nuclear University MEPhI (Moscow Engineering Physics Institute), Moscow, 115409 Russia}
\author{L.\,Zhou}
\affiliation{Innovation Academy for Precision Measurement Science and Technology, Chinese Academy of Sciences, Wuhan 430071, China}
\author{M.\,S.\,Zhan}
\affiliation{Innovation Academy for Precision Measurement Science and Technology, Chinese Academy of Sciences, Wuhan 430071, China}


\begin{abstract}
We investigate the shift caused by asymmetry of spectroscopic lineshape in atomic interferometers, which has not previously been discussed in the scientific literature. This asymmetry arises because laser field is frequency-chirped not only during the free-evolution intervals of atoms, but also during the Ramsey pulses. As a result, the effective detuning from the working atomic transition during the pulses also depends on the chirping rate, which, in turn, leads to the lineshape-asymmetry-caused shift (LACS). It is shown that this shift has an inverse cubic dependence of $\propto 1/T^3$ on the duration of the interval between the Ramsey pulses $T$, which markedly contrasts with the $\propto 1/T^2$ dependence typical in atomic interferometry. Therefore, the metrological importance of this shift substantially increases for compact atomic interferometers with a short baseline. For example, for interferometers-gravimeters using two-photon transitions in rubidium atoms, at $T\sim 1$~ms we estimate the LACS shift and its variations at the level of 0.1-1~mGal, while for $T\sim 100$~$\mu$s this can reach a value of 0.1-1~Gal.
\end{abstract}

\maketitle
Atomic interferometry is the basis of modern quantum sensors for various purposes. For example, quantum absolute gravimeters based on the interference of ultracold atoms were first implemented in 1991 \cite{Kasevich_1991} and in the next demonstrated outstanding results in the field of high-precision measurements of the free-fall acceleration \cite{Peters_2001,Freier_2016,Menoret_2018,Heine_2020}. Their high sensitivity, long-term stability, and lack of need for calibration make them an indispensable tool for a wide range of fundamental \cite{Wicht_2002,Rosi_2014,Duan_2016,Asenbaum_2020,Zhou_2022} and practical applications \cite{Kennedy_2016,Diament_2024}.

Current trends in the development of atomic interferometers are aimed at miniaturization and increased mobility, which inevitably leads to a reduction of the transit time of atoms in the interferometer. For example, in high-sensitive transportable quantum gravimeters, the interferometric cycle time can reach several hundred milliseconds \cite{Freier_2016}. This, in turn, significantly limits the compactness, response time, and dynamic range of the sensor, and increases its sensitivity to sources of phase noise and systematic shifts such as low-frequency vibration, residual magnetic field gradients, spatial inhomogeneity of the intensity and phase of light fields, etc.

In this context, short-baseline atomic interferometers (i.e., with small atomic flight time) represent a promising alternative. They exhibit significantly lower sensitivity to external disturbances, high sampling rates (tens of hertz), and great potential for miniaturization. Moreover, as shown in \cite{McGuinness_2012, Narducci_2022}, even with interferometric atomic sampling sequences on the order of milliseconds, it is possible to achieve metrological characteristics that meet the requirements of most practical applications. This is facilitated by the development of new approaches to increasing the phase stability and contrast of interference signals \cite{Olivares_2020, Qing_2023, Wang_2024}, as well as effective methods for analyzing, correcting, and processing measurements \cite{Cheiney_2018, Wei_2025}. Thus, short-baseline interferometers represent a promising direction for the development of high-precision compact quantum sensors.

In this paper, we present a theoretical study of the systematic shift caused by the asymmetry of the signal lineshape in atomic interferometers, which has not previously been discussed in the scientific literature. In particulary, we demonstrate that for short-baseline interferometer-gravimeters, this shift can be dominant, significantly limiting the accuracy and stability of compact devices.

As a specific scheme of an atomic interferometer, we will consider a quantum gravimeter using an atomic transition $|g\rangle$$\leftrightarrow$$|e\rangle$ with an unperturbed frequency $\omega_0$, where $|g\rangle$ and $|e\rangle$ are the ground and excited states, respectively, which correspond to the following column vectors:
\begin{equation}\label{basis}
|g\rangle=\left(
            \begin{array}{c}
              0 \\
              1 \\
            \end{array}
          \right),\quad
|e\rangle=\left(
            \begin{array}{c}
              1 \\
              0 \\
            \end{array}
             \right).
\end{equation}
In this two-level atom, transitions between states $|g\rangle$ and $|e\rangle$ during Ramsey pulses are induced under the action of a resonant interaction operator of the plane running wave type with the wave vector $\textbf{k}_{\rm eff}$:
\begin{equation}\label{wave}
\widehat{V}(t,\textbf{r})=-\frac{\hbar\Omega_0}{2} e^{-i[\omega t+\alpha t^2/2-(\textbf{r}\cdot\textbf{k}_{\rm eff})+\phi]}|e\rangle\langle g|+h.c.\,,
\end{equation}
where $\omega$ is the carrier frequency, which can vary near the transition frequency $\omega_0$, $\alpha$ is the speed of the additional frequency chirp, $\phi$ is some fixed phase, and $\Omega_0>0$ is the real Rabi frequency of the induced transitions. As is known, in such an interferometric scheme, the quantity $(\textbf{g}\cdot \textbf{k}_{\rm eff})$ can be measured, where $\textbf{g}$ is the free fall acceleration due to gravity.

Note that the interaction type (\ref{wave}) occurs both for one-photon transition under the action of a running light wave in the case of the optical transition $|g\rangle$$\leftrightarrow$$|e\rangle$, and for two-photon transition in the bichromatic field of two counterpropagating running light waves, when the difference in the optical frequencies of these waves ($\omega_1-\omega_2$) varies near the microwave frequency $\omega_0$ in the case of the transition $|g\rangle$$\leftrightarrow$$|e\rangle$ between hyperfine levels of an atom in the ground state (for example, in alkali metal atoms). However, this does not matter for our work.

\begin{figure}[t]
    \includegraphics[width=0.9\linewidth]{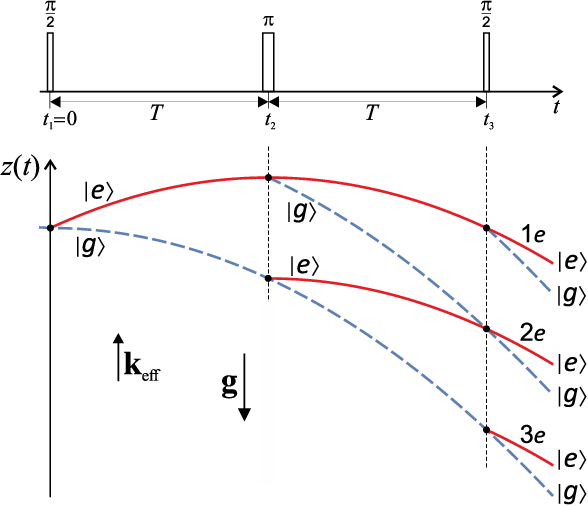}
    \caption{Standard scheme of a Mach–Zehnder type atomic interferometer gravimeter.}\label{Rams_pulses}
\end{figure}

As an example, we will consider a standard Mach-Zehnder atomic interferometer-gravimeter based on a sequence of three Ramsey pulses ($\pi/2$--$\pi$--$\pi/2$ sequence), shown in Fig.\,\ref{Rams_pulses}. The general theory of atomic interferometry is described in \cite{Storey_1994,Borde_2001}.

In the reference frame of a moving atom, the operator (\ref{wave}) can be represented as
\begin{equation}\label{atomic_ref}
\widehat{V}^{\rm (atom)}=-\frac{\hbar\Omega_0}{2} e^{-i\varphi(t)}|e\rangle\langle g|+h.c.\,,
\end{equation}
with a time-dependent phase, $\varphi(t)$. Then, for sufficiently short durations of Ramsey pulses, the evolution operator of an atom under the action of the $j$-th Ramsey pulse can be written as
\begin{equation}\label{j_evolution}
\widehat{R}^{(j)}=\left(
    \begin{array}{cc}
      A^{}_j & iB_j e^{-i\varphi(t_j)} \\
      iB_j e^{i\varphi(t_j)} & A^{\ast}_j \\
    \end{array}
  \right),
\end{equation}
where $t_j$ is the point in time of the $j$-th Ramsey pulse (see Fig.\,\ref{Rams_pulses}). The matrix elements in (\ref{j_evolution}) are defined as follows
\begin{eqnarray}\label{matrix_el}
&&A_j= \cos\bigg(\frac{\Omega^{(j)}\tau_j}{2}\bigg)+i\frac{\delta_j}{\Omega^{(j)}}\sin\bigg(\frac{\Omega^{(j)}\tau_j}{2}\bigg),\nonumber\\
&&B_j=\frac{\Omega_0}{\Omega^{(j)}}\sin\bigg(\frac{\Omega^{(j)}\tau_j}{2}\bigg),\\
&& \Omega^{(j)}=\sqrt{\Omega^2_0+\delta^2_j}\,,\nonumber
\end{eqnarray}
where $\tau_j$ is the duration of the $j$-th pulse, $\delta_j$ is the effective detuning from the resonance during the $j$-th pulse from the point of view of the moving atom, $\Omega^{(j)}$ is the generalized Rabi frequency during the $j$-th pulse. Note that $B_j$ is a real number.

Let us consider the usual scheme of interrogating atoms in the case of spatially uniform acceleration of gravity $\textbf{g}$. Let at time $t=0$ (i.e., at the beginning of the first Ramsey pulse, see Fig.\,\ref{Rams_pulses}) the atoms are in the lower state $|g\rangle$ and have an initial velocity $\textbf{v}$. Then, as the spectroscopic signal after the action of the Ramsey sequence in Fig.\,\ref{Rams_pulses}, we will consider the total population $N_e$ of atoms in the upper state $|e\rangle$ on three different trajectories: 1$e$, 2$e$ and 3$e$ (see the red solid lines in Fig.\,\ref{Rams_pulses}). We assume that after the action of three Ramsey pulses all three trajectories have diverged by distances significantly greater than the de Broglie wavelength. In this case, we have the following expression
\begin{equation}\label{N_e}
N_e=n_{1e}+n_{2e}+n_{3e}=|K_{1e}|^2+|K_{2e}|^2+|K_{3e}|^2.
\end{equation}
Using the notation from (\ref{j_evolution}), we can write
\begin{eqnarray}\label{K}
  K_{1e} &=&B_1 A_2 A_3,\nonumber  \\
  K_{2e} &=&B_1 B_2 B_3- A^{\ast}_1B_2 A_3 e^{i\widetilde{\alpha}T^2}, \\
  K_{3e} &=& A^{\ast}_1 A^{\ast}_2 B_3,\nonumber
\end{eqnarray}
where $T$ is the time between pulses (see Fig.\,\ref{Rams_pulses}), and $\widetilde{\alpha}$ is the frequency chirp speed centered on the measured value $(\textbf{g}\cdot \textbf{k}_{\rm eff})$
\begin{equation}\label{alpha}
\widetilde{\alpha}=\alpha - (\textbf{g}\cdot \textbf{k}_{\rm eff}).
\end{equation}
In the standard interrogation scheme Fig.\,\ref{Rams_pulses}, we have the following expressions for the duration and  detuning of each pulse
\begin{eqnarray}\label{delta_j}
&& \tau_1=\tau_3=\tau,\quad \tau_2=2\tau, \nonumber\\
&& \delta_1=\Delta_{\rm D},\quad \delta_2= \widetilde{\alpha}T+\Delta_{\rm D},\quad  \delta_3= 2\widetilde{\alpha}T+\Delta_{\rm D},
\end{eqnarray}
where $\Delta_{\rm D}$ is the general residual contribution to the detuning from resonance, which, taking into account the recoil effect for atoms with mass $M$, has the following form
\begin{equation}\label{Delta}
\Delta_{\rm D}=\omega-\omega_0-\frac{\hbar|\textbf{k}_{\rm eff}|^2}{2M}-(\textbf{v}\cdot \textbf{k}_{\rm eff})-\overline{\delta}_{\rm ac}\,,
\end{equation}
where $-(\textbf{v}\cdot \textbf{k}_{\rm eff})$ is the Doppler shift during the action of the first Ramsey pulse at $t=0$ (see Fig.\,\ref{Rams_pulses}). In (\ref{Delta}) we also take into account the ac Stark shift $\overline{\delta}_{\rm ac}$ of the atomic transition frequency caused by the probe light field, which we assume to be the same for all three pulses.

\begin{figure}[t]
    \includegraphics[width=0.7\linewidth]{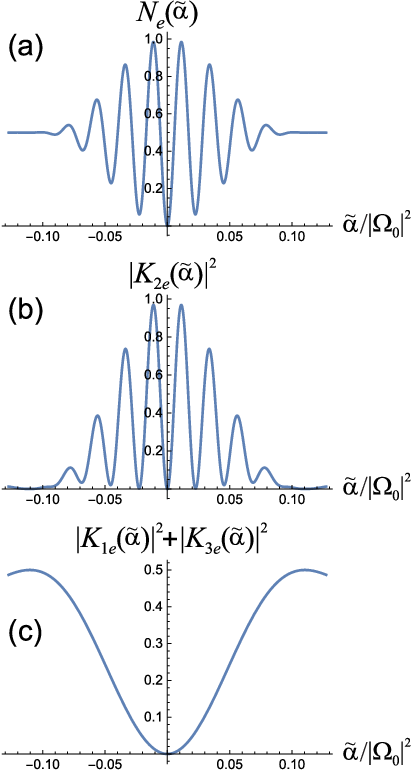}
    \caption{Typical forms of dependencies: (a)\;total signal $N_e(\widetilde{\alpha})$; (b)\;interference contribution $|K_{2e}(\widetilde{\alpha})|^2$; (c)\;substrate ($|K_{1e}(\widetilde{\alpha})|^2+|K_{3e}(\widetilde{\alpha})|^2$). Calculation parameters: $\tau\Omega_0=\pi/2$, $T=10\tau$, $\Delta_{\rm D}=0$.}\label{lineshape}
\end{figure}

Let us now investigate the functional dependence of the signal $N_e(\widetilde{\alpha})$ on the variable parameter $\widetilde{\alpha}$
\begin{equation}\label{N_Rams}
N_e(\widetilde{\alpha})=|K_{1e}(\widetilde{\alpha})|^2+|K_{2e}(\widetilde{\alpha})|^2+|K_{3e}(\widetilde{\alpha})|^2.
\end{equation}
In general, the dependence $N_e(\widetilde{\alpha})$ on the variable $\widetilde{\alpha}$ is a sequence of narrow Ramsey resonances with a width of about $\pi/T^2$ and a smooth envelope (the width is of the order of $\sim 2\Omega_0/T$), which is described by the interference contribution $|K_{2e}(\widetilde{\alpha})|^2$ in (\ref{N_Rams}). In addition, there is also a wide smooth dependence (the width is of the order of $\sim 2\Omega_0/T$), which is formed by two other terms $|K_{1e}(\widetilde{\alpha})|^2$ and $|K_{3e}(\widetilde{\alpha})|^2$. The typical form of the corresponding dependences is shown in Fig.\,\ref{lineshape}.

\begin{figure}[t]
    \includegraphics[width=0.8\linewidth]{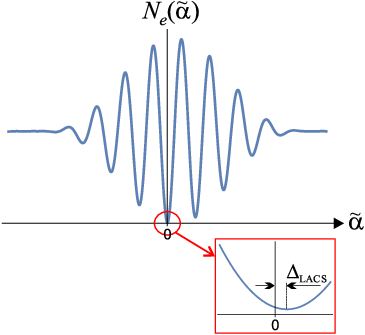}
    \caption{Demonstration of the shift in the position of the extremum at the central Ramsey resonance in the dependence $N_e(\widetilde{\alpha})$ under $\Delta_{\rm D}\neq 0$.}\label{shift}
\end{figure}

\begin{figure}[t]
    \includegraphics[width=0.8\linewidth]{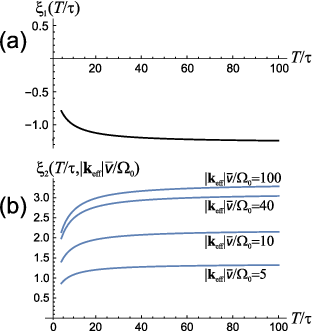}
    \caption{(a)\,Dependence of the dimensionless function $\xi^{}_1(T/\tau)$ in the expression (\ref{shift_dep}) at $T\gg \tau$. (b)\,Dependence of the dimensionless function $\xi^{}_2(T/\tau,|{\bf k}_{\rm eff}|\bar{v}/\Omega_0)$ on $T/\tau$ in the expression (\ref{LACS_v}) for several fixed values of $|{\bf k}_{\rm eff}|\bar{v}/\Omega_0$ (indicated in the plot).}
    \label{fT}
\end{figure}

In this case, the gravimetric measurement of the value $(\textbf{g}\cdot \textbf{k}_{\rm eff})$ corresponds to the position $\widetilde{\alpha}_{\rm ext}$ of the central Ramsey resonance peak near zero (see Fig.\,\ref{shift}). The $\widetilde{\alpha}_{\rm ext}$ value itself determines the shift of the gravimetric measurement. Based on the expressions (\ref{matrix_el}), (\ref{K}) and (\ref{delta_j}), it is easy to show that in the case of $\Delta_{\rm D}=0$ the following relations are satisfied: $K_{1e}(-\widetilde{\alpha})=K_{1e}^*(\widetilde{\alpha})$, $K_{2e}(-\widetilde{\alpha})=K_{2e}^*(\widetilde{\alpha})$ and $K_{3e}(-\widetilde{\alpha})=K_{3e}^*(\widetilde{\alpha})$ [because of $\delta_j(-\widetilde{\alpha})$=$-\delta_j(\widetilde{\alpha})$]. However, for $\Delta_{\rm D}\neq 0$ these relations are violated: $K_{1e}(-\widetilde{\alpha})\neq K_{1e}^*(\widetilde{\alpha})$, $K_{2e}(-\widetilde{\alpha})\neq K_{2e}^*(\widetilde{\alpha})$ and $K_{3e}(-\widetilde{\alpha})\neq K_{3e}^*(\widetilde{\alpha})$ [because of $\delta_j(-\widetilde{\alpha})$$\neq$$-\delta_j(\widetilde{\alpha})$]. As a result, we have the following relations
\begin{eqnarray}\label{simmetry}
&& N_e(-\widetilde{\alpha})=N_e(\widetilde{\alpha})\quad {\rm for}\quad \Delta_{\rm D}=0,\nonumber \\
&& N_e(-\widetilde{\alpha})\neq N_e(\widetilde{\alpha})\quad {\rm for}\quad \Delta_{\rm D}\neq 0.
\end{eqnarray}
Thus, for $\Delta_{\rm D}=0$ the signal $N_e(\widetilde{\alpha})$ is an even (symmetric) function on $\widetilde{\alpha}$, and, therefore, the position of the local extremum at the central Ramsey resonance is unshifted, i.e. $\widetilde{\alpha}_{\rm ext}=0$. However, for $\Delta_{\rm D}\neq 0$, the signal $N_e(\widetilde{\alpha})$ is no longer a symmetric function on $\widetilde{\alpha}$, which means, in our case, a shift in the position of the central Ramsey resonance peak: $\Delta^{}_{\rm LACS}\neq 0$ (see Fig.\,\ref{shift}). This can be characterized as a shift caused by the lineshape asymmetry (lineshape-asymmetry-caused shift, LACS). Calculations show that in the case of $\Omega_0\tau=\pi/2$ and $|\Delta_{\rm D}|<0.5\Omega_0$, the following linear dependence on $\Delta_{\rm D}$ is well satisfied:
\begin{equation}\label{shift_dep}
\Delta^{}_{\rm LACS}\approx \xi^{}_1(T/\tau)\frac{\tau}{T^3}\frac{\Delta_{\rm D}}{\Omega_0}\,,
\end{equation}
where the dimensionless function $\xi^{}_1(T/\tau)<0$ depends rather slowly on the ratio $T/\tau$ in the case of $T\gg \tau$ (see Fig.\,\ref{fT}). Thus, the dependence of the detunings $\delta_j(\widetilde{\alpha})$ on the parameter $\widetilde{\alpha}$ [see expressions in (\ref{delta_j})] and the presence of general residual contribution $\Delta_{\rm D}\neq 0$ to the detunings from the resonance during the action of the Ramsez pulses leads to a shift $\Delta^{}_{\rm LACS}\neq 0$ when measuring the quantity $(\textbf{g}\cdot \textbf{k}_{\rm eff})$. It should be noted that this shift is not a phase shift in the standard meaning, but is the result of asymmetry in both the envelope of the interference term $|K_{2e}(\widetilde{\alpha})|^2$ and the substrate ($|K_{1e}(\widetilde{\alpha})|^2+|K_{3e}(\widetilde{\alpha})|^2$).

In the case of a non-monovelocity atomic ensemble, it is necessary to consider the signal averaged over the velocity distribution $f({\bf v})$:
\begin{equation}\label{Ne_v}
\langle N_e(\widetilde{\alpha})\rangle_{\bf v}=\int f({\bf v})N_e(\widetilde{\alpha})d{\bf v}^3,
\end{equation}
and the LACS value will undergo significant changes relative to the expression (\ref{shift_dep}). For example, consider an atomic ensemble with a Maxwellian velocity distribution
\begin{equation}\label{fv_M}
 f_{\rm M}({\bf v})=\frac{1}{\pi^{3/2}\bar{v}^3}e^{-\frac{({\bf v}-{\bf v}_0)^2}{\bar{v}^2}},
\end{equation}
where $\bar{v}$ is the width parameter of the velocity distribution relative to the average velocity of the atomic beam ${\bf v}_0$. In the general case, there is no simple expression for LACS. However, for $\Omega_0\tau=\pi/2$ and $T\gg \tau$, in the case where the Doppler width $|{\bf k}_{\rm eff}| \bar{v}$ significantly exceeds the Rabi frequency, $|{\bf k}_{\rm eff}| \bar{v}>5 \Omega_0$, the following expression works quite well
\begin{equation}\label{LACS_v}
\Delta^{\langle {\bf v}\rangle}_{\rm LACS}\approx \xi^{}_2(T/\tau,|{\bf k}_{\rm eff}|\bar{v}/\Omega_0) \frac{\tau}{T^3}\frac{\Omega_0}{|{\bf k}_{\rm eff}| \bar{v}}\frac{\langle\Delta_{\rm D}\rangle_{\bf v}}{|{\bf k}_{\rm eff}| \bar{v}}\,,
\end{equation}
where $\langle\Delta_{\rm D}\rangle_{\bf v}$ is the residual Doppler detuning for the average atomic velocity $\textbf{v}_0$:
\begin{equation}\label{Delta_v}
\langle\Delta_{\rm D}\rangle_{\bf v}=\omega-\omega_0-\frac{\hbar|\textbf{k}_{\rm eff}|^2}{2M}-(\textbf{v}_0\cdot \textbf{k}_{\rm eff})-\overline{\delta}_{\rm ac}\,,
\end{equation}
and $\xi^{}_2(T/\tau,|{\bf k}_{\rm eff}|\bar{v}/\Omega_0)$ is a dimensionless function that depends rather slowly on two dimensionless parameters $T/\tau$ and $|{\bf k}_{\rm eff}|\bar{v}/\Omega_0$ [see Fig.\,\ref{fT}(b)]. Formula (\ref{LACS_v}) is well satisfied for $|\langle\Delta_{\rm D}\rangle_{\bf v}|<0.5|{\bf k}_{\rm eff}| \bar{v}$. Comparing the signs of the dependences in Fig.\,\ref{fT}(a) and (b), it is interesting to note that the shift (\ref{LACS_v}) is opposite in sign to the shift for a mono-velocity beam (\ref{shift_dep}).

Let us estimate the metrological significance of the LACS using the example of atomic gravimeter-interferometer. Obviously, due to the inverse cubic dependence ($\propto 1/T^3$), the role of this shift in the error budget will increase for compact interferometers with a short baseline. Consider, for example, the following parameters of the Ramsey atomic interrogation scheme: $T=2.5$~ms, $\tau=25$~$\mu$s, and $\Omega_0=2\pi\times 10$~kHz (i.e. $\Omega_0 \tau=\pi/2$). With regard to the velocity distribution of the atomic beam, we use the following values: ${\bf v}_0\sim 50$~cm/s and $\bar{v}\sim 5$~cm/s, which for two-photon transition through the D lines in rubidium atoms gives estimates of $|({\bf k}_{\rm eff}\cdot {\bf v}_0)|\sim 1$~MHz and $|{\bf k}_{\rm eff}| \bar{v}\sim 100$~kHz. In this case, taking into account the possibility of a one-percent variation in the value of $|{\bf v}_0|$ during gravimetric measurements, we can safely assume an estimate of $|\langle\Delta_{\rm D}\rangle_{\bf v}|\gtrsim 10$~kHz both for the value of $|\langle\Delta_{\rm D}\rangle_{\bf v}|$ itself and for its variations. Substituting these values into the formula (\ref{LACS_v}), we obtain an estimate of the LACS value and its variations at the level of $|\Delta^{\langle {\bf v}\rangle}_{\rm LACS}|\gtrsim 20$~s$^{-2}$, which, when applied to gravimetry, corresponds to $\gtrsim 0.1$~mGal. Thus, on the millisecond scale of free evolution times in atomic interferometer-gravimeters, the LACS shift can lead to measurement errors and fluctuations at the level of 0.1-1~mGal. Accordingly, for a free evolution time $T\sim 100$~$\mu$s, this estimate reaches values of 0.1-1~Gal. Obviously, shifts of this magnitude can significantly reduce the metrological characteristics of short-baseline atomic interferometers. Moreover, in the case of a mono-velocity atomic beam, this shift can reach even greater values. We also note that although for interferometer-gravimeters with a free evolution time at the level of tens of milliseconds the LACS shift is not so critically large, it is nevertheless not negligibly small (it can reach the order of 1-10~$\mu$Gal) and it should also be taken into account in the measurement error budget.

For the sake of general discussion, it should be noted that the most compact quantum gravimeters owe their efficiency to a high interrogation rate of atoms (tens of hertz) and the recapturing of cold atoms in the vapor cell into a magneto-optical trap at the end of the experimental cycle. This, in turn, sets the total free-fall time of atoms at 40-50 ms (8-12 mm), determined by the size of the laser beam overlap region, the recapture efficiency, the required response time, etc. During the free-fall time of atoms, three stages occur: preparation of the initial atomic wave packet, atomic interference, and detection. The duration of the atom preparation stage depends on the preparation method, but averages several tens of milliseconds. The duration of the detection stage after the Ramsey sequence is typically several milliseconds \cite{Peters_2001}. The remaining time (approximately 20 ms or less) is accounted for by the atomic interference stage itself. An example of a compact gravimeter was demonstrated in \cite{McGuinness_2012}, where the $T$ value did not exceed 7~ms, and the sensitivity reached from 570~$\mu$Gal/$\sqrt{Hz}$ to 36.7~mGal/$\sqrt{Hz}$ at interrogation rates from 50~Hz to 330~Hz, respectively. It should be noted that gravimeters with a short time $T$ have a high response time, increased stability to many systematic shifts (e.g., magnetic field gradient, spatial non-uniformity of intensity and wavefront curvature, etc.) and to vibrations, making it possible to use quantum gravimeters in highly dynamic conditions and on moving platforms. All of the above, taking into account the general trend of device miniaturization, demonstrates the relevance of our results for atomic gravimeters, in which the interval between Ramsey pulses $T$ does not exceed 10-20~ms, and especially at $T$ of the order of milliseconds or less. Moreover, in addition to quantum gravimeters, one can also consider other devices (for example, atomic accelerometers and gyroscopes) that use chirping of the probe laser frequency.

In conclusion, we theoretically investigated the shift in atomic interferometers, which has not previously been discussed in the scientific literature. This shift is caused by the lineshape asymmetry of the spectroscopic signal (LACS) when using the frequency-chirped probe field method and has an inverse cubic dependence of $\propto 1/T^3$ on the duration of the interval between the Ramsey pulses $T$, which contrasts markedly with the typical dependence for atomic interferometry $\propto 1/T^2$. Thus, its influence increases significantly for atomic interferometers with a short baseline. Note that this shift is not a phase shift in the standard meaning, but is the result of asymmetry in both the envelope of the interference term and the substrate. Using the example of a Mach-Zehnder type interferometer-gravimeter, it was shown that even starting with $T$ of tens milliseconds, the LACS shift can have metrological significance, and at $T$ of milliseconds and below, it can become dominant, substantially limiting the accuracy and stability of small-sized devices. Therefore, its experimental detection, detailed study, and development of suppression methods are relevant tasks for atomic interferometry and the creation of precision compact and mobile quantum sensors for various purposes (gravimeters, accelerometers, and gyroscopes).


\begin{thebibliography}{99}
%
\bibitem{Kasevich_1991}
M. Kasevich, S. Chu, Phys. Rev. Lett. \textbf{67}, 181 (1991).
%
\bibitem{Peters_2001}
A. Peters, K. Y. Chung, S. Chu, Metrologia \textbf{38}, 25 (2001).
%
\bibitem{Freier_2016}
C. Freier, M. Hauth, V. Schkolnik, et al., J. Phys.: Conf. Ser. \textbf{723}, 012050 (2016).
%
\bibitem{Menoret_2018}
V. M\'{e}noret, P. Vermeulen, N. Le Moigne, et al., Sci. Rep. \textbf{8}, 12300 (2018).
%
\bibitem{Heine_2020}
N. Heine, J. Matthias, M. Sahelgozin, et al., Eur. Phys. J. D \textbf{74}, 174 (2020).
%
\bibitem{Wicht_2002}
A. Wicht, J. M. Hensley, E. Sarajlic, et al., Phys. Scr. \textbf{102}, 82 (2002).
%
\bibitem{Rosi_2014}
G. Rosi, F. Sorrentino, L. Cacciapuoti, et al., Nature \textbf{510}, 518 (2014).
%
\bibitem{Duan_2016}
X. C. Duan, X. B. Deng, M. K. Zhou, et al., Phys. Rev. Lett. \textbf{117}, 023001 (2016).
%
\bibitem{Asenbaum_2020}
P. Asenbaum, C. Overstreet, M. Kim, et al., Phys. Rev. Lett. \textbf{125}, 191101 (2020).
%
\bibitem{Zhou_2022}
L. Zhou, S.-T. Yan, Y.-H. Ji, et al., Toward a high-precision mass-energy test of the equivalence principle with atom interferometers, Front. Phys. \textbf{10}, 1039119 (2022).
%
\bibitem{Kennedy_2016}
J. Kennedy, T. P. A. Ferr\'{e}, B. Creutzfeldt, Water Resources Research \textbf{52}, 7244-7261 (2016).
%
\bibitem{Diament_2024}
M. Diament, G. Lion, G. Pajot-M\'{e}tivier, et al., IEEE Instrum. Meas. Mag. \textbf{27}, 17 (2024).
%
\bibitem{McGuinness_2012}
H. J. McGuinness, A. V. Rakholia, G. W. Biedermann, Appl. Phys. Lett. \textbf{100}, 011106 (2012).
%
\bibitem{Narducci_2022}
F. A. Narducci, A. T. Black, J. H. Burke, Adv. Phys.: X \textbf{7}, 1946426 (2022).
%
\bibitem{Olivares_2020}
G. A. Olivares-Renter\'{\i}a, D. A. Lancheros-Naranjo, E. Gomez, et al., Phys. Rev. A \textbf{101}, 043613 (2020).
%
\bibitem{Qing_2023}
H. Qing-Qing, Z. Hang, L. Yu-Kun, et al., Optik \textbf{276}, 170637 (2023).
%
\bibitem{Wang_2024}
Y. Wang, J. Glick, T. Deshpande, et al., Phys. Rev. Lett. \textbf{133}, 243403 (2024).
%
\bibitem{Cheiney_2018}
P. Cheiney, L. Fouch\'{e}, S. Templier, et al., Phys. Rev. Applied \textbf{10}, 034030 (2018).
%
\bibitem{Wei_2025}
J. Wei, J. Huang, C. Lee, Phys. Rev. Res. \textbf{7}, L012064 (2025).
%
\bibitem{Storey_1994}
P. Storey, C. Cohen-Tannoudji, The Feynman path integral approach to atomic interferometry. A tutorial. J. Phys. II France \textbf{4}, 1999-2027 (1994).
%
\bibitem{Borde_2001}
C. J. Bord\'{e}, Theoretical tools for atom optics and interferometry, C. R. Acad. Sci. Paris \textbf{2}, S\'{e}rie IV, 509-530 (2001).
%
%
\bibitem{McGuinness_2012}
H. J. McGuinness, A. V. Rakholia, G. W. Biedermann, Appl. Phys. Lett. \textbf{100}, 011106 (2012).
%
%


\end{thebibliography}
\end{document}